\documentclass[pre]{revtex4-1}
\usepackage{amssymb}
\usepackage{graphicx}
\usepackage{amsmath}
\usepackage{amssymb}

\begin{document}
\preprint{}

\title{Inference and learning in sparse systems with multiple states}

\author{A. Braunstein}
\email{alfredo.braunstein@polito.it}
\affiliation{Human Genetics Foundation, Via Nizza 52, 10126 Torino, Italy}
\affiliation{Politecnico di Torino, C.so Duca degli Abruzzi 24, 10129 Torino, Italy}

\author{A. Ramezanpour,}
\email{abolfazl.ramezanpour@polito.it}
\affiliation{Politecnico di Torino, C.so Duca degli Abruzzi 24, 10129 Torino, Italy}

\author{R. Zecchina}
\email{riccardo.zecchina@polito.it}
\affiliation{Politecnico di Torino, C.so Duca degli Abruzzi 24, 10129 Torino, Italy}
\affiliation{Human Genetics Foundation, Via Nizza 52, 10126 Torino, Italy}
\affiliation{Collegio Carlo Alberto, Via Real Collegio 30, 10024 Moncalieri, Italy}

\author{P. Zhang}
\email{pan.zhang@polito.it}
\affiliation{Politecnico di Torino, C.so Duca degli Abruzzi 24, I-10129 Torino, Italy}

\date{\today}
\begin{abstract}

We discuss  how inference can be performed when data are  sampled from the non-ergodic phase of systems with multiple attractors.
 We take as model system the  finite connectivity Hopfield model in the memory phase and suggest a cavity method approach to reconstruct the couplings when the data are separately sampled from  few attractor states.
 We also show  how the  inference results can be converted into  a learning protocol for neural networks in which patterns are presented through weak external fields. The protocol is simple and fully local, and is able to store patterns with a finite overlap with the input patterns without ever reaching a spin glass phase where all memories are lost.

\end{abstract}
 \maketitle
\section{Introduction}\label{S1}
The  problem  of inferring interactions couplings in complex systems  arises from the huge quantity of empirical data that are being made available in many fields of science and from the difficulty of making systematic measurements on interactions. In biology, for example, empirical data on neurons populations,  small molecules, proteins and genetic interactions, have largely outgrown the understanding of the underlying system mechanisms. In all these cases the inverse problem, whose aim is to infer some effective  model from the empirical data with just partial a priori knowledge, is of course extremely relevant.

Statistical physics,  with its set of well-understood theoretical models, has the crucial role to provide complex, but clear-cut benchmarks, whose direct solution is known and that can therefore be used to develop and test new inference methods.

In a nutshell, the equilibrium approach to inverse problems consists inferring some information about  a system defined through an  energy function $H(\underline{\sigma})$  starting from a set of sampled equilibrium configurations.
Suppose  $M$ i.i.d sampled configurations $\mathcal{S} = \{\underline{\sigma}^{(i)}\}_{i=1\dots M}$ generated by a Boltzmann distribution of an unknown energy function  $H$, $\mathcal{P}(\underline{\sigma})=\frac{1}{Z_H}e^{-H(\underline{\sigma})}$  are given.
The posterior distribution of $H$ (also called likelihood) is given by $P(H|\mathcal{S})\propto \exp(-M(\left<H\right> + \log Z_H)) P(H)$,
where $\left<H\right>=\frac{1}{M}\sum_{i=1}^M H(\underline{\sigma}^{(i)})$ represents the average of $H$ over the given sample configurations and $P(H)$ the prior knowledge about $H$. The parameter $M$ plays the role of an inverse temperature: when $M$ is very large, $P(H|\mathcal{S})$ peaks on the maximums (with respect to $H$) of the `log-likelihood` $\mathcal{L} = -\left<H\right> - \log Z_H$. The problem of identifying the maximum of $\mathcal L$  can be thought of as an optimization problem, normally very difficult both because the space of $H$ is large and because $\log Z_H$ is very difficult to estimate by itself on a candidate solution.

Several methodological advances and  stimulating  preliminary applications in
neuroscience have been put forward in the last few years \cite{Bialek-nature-2006,Bialek-arxiv-2007,Monasson-jphysa-2009,Monasson-pnas-2009}. Still the
field presents several major conceptual and methodological open problems
related to both the efficiency and the accuracy of the methods. One  problem
which we consider here is how to perform inference when data are not coming
from a uniform sampling over the equilibrium configurations of a system but
rather they are taken from a subset of all the attractive states.  This case
arises for instance when  we consider systems with multiple attractors and we
want to reconstruct the interactions couplings from  measurements coming from a
subset of the attractors.

In what follows, we take as model system the Hopfield model over random graphs
in its memory phase (i.e. with multiple equilibrium states) and show how the
interaction couplings can be inferred from data taken from a subset of memories
(states). This will be done by employing the Bethe equations (normally used in 
the ergodic phase where they are asymptotically exact) by taking advantage of a certain property of their multiple fixed points in the non-ergodic phase. The method can be used to infer both couplings and external local fields.

We also show how from the inference method one can derive a simple
unsupervised learning protocol which is able to learn patterns in presence of
week and highly fluctuating input signals, without ever reaching a spin glass
like saturation regime in which all the memories are lost. 
The technique that we will discuss  is based on the so called cavity method and leads to a distributive algorithmic implementation generically known as  message--passing scheme.

The paper is organized as follows. First, in Section \ref{S2} we define the problem, and make connections with related works. Section \ref{S3} is  concerned with the  inference problem in non-ergodic regimes, for which a simple algorithmic approach is presented.
In Section \ref{S4} we apply the technique to the finite connectivity Hopfield model in the memory phase.
Section \ref{S5} shows how the approach can be turned into an unsupervised learning protocol.
Conclusions and perspectives are given in Section \ref{S6}.

\section{The inverse Hopfield problem}\label{S2}

The Hopfield model is a simple neural network model with pair-wise interactions which behaves as an  attractor associative memory \cite{Hopfield-pnas-1982}.
Its phase diagram  is known exactly when memories are random patterns and the model is defined over either fully connected or sparse graphs \cite{Amit-prl-1985,Coolen-jphysa-2003}. Reconstructing interactions  in the Hopfield model  from partial data thus represents a natural benchmark problem for tools which pretend to  be applied to data coming from  multi electrode measurements from large collections of neurons.  The underlying idea is that  a statistically consistent interacting model (like the Hopfield model)  inferred from the data could  capture some aspects of the system which are not  easy to grasp from the raw data \cite{Bialek-nature-2006}.  Here we limit our analysis to artificial data.

In the Hopfield model the couplings $J_{ij}$  are given by the covariance matrix of a set of random patterns which represent the memories to be stored in the system. We will use the model to generate data through sampling and we will aim at inferring the couplings.

The structure of the phase space of the Hopfield model  at sufficiently low temperature and for a not too large number of patterns is divided into  clusters of configurations which are highly correlated with the patterns.  We will proceed by sampling configurations from a subset of clusters and try to infer the interactions.
The simple observation that  we want to exploit is the fact that fluctuations within single clusters are heavily influenced by the existence of other clusters and thus contain information about the total system.

We consider a system of $N$ binary neurons $\sigma_i \in \{-1,+1\}$ (or Ising spins) interacting over a random regular graph of degree $K$;
that is every node has a fixed number $K$ of neighbors which are selected randomly.  
The connectivity pattern is defined by the elements $a_{ij}\in \{0,1\}$ of the adjacency matrix.
The (symmetric) interactions between two neighboring neurons are given by the Hebb rule, i.e. $J_{ij}=J_{ji}=\frac{a_{ij}}{K} \sum_{\mu=1}^P \xi_i^{\mu} \xi_j^{\mu}$,
where  $\underline{\xi}^{\mu}$ are the patterns to be memorized and $P$ is their number.  At finite temperature $T=1/\beta$, we simulate the system by a  Glauber dynamics  \cite{Glauber-jmathphys-1963};
starting from an initial configuration, the spins are chosen in a random sequential way and flipped with the following transition probability 
\begin{equation}
W(\sigma_i\to-\sigma_i)=\frac{1-\sigma_i\tanh\beta h_i}{2}.
\end{equation}
where
\begin{equation}
h_i=\theta_i + \sum_{j \in \partial i} J_{ij}\sigma_j,
\end{equation}
is the local field experienced by spin $i$ and $\theta_i$ is an external field. We use $\partial i$ to denote the set of neighbors interacting with $i$. 
The process satisfies detailed balance and at  equilibrium  the probability of steady state configurations $\underline{\sigma}$ is given by the Gibbs measure
\begin{equation}\label{gibbs}
\mathcal{P}(\underline{\sigma})=\frac{1}{Z[\underline{J},\underline{\theta}]}e^{\beta \sum_i \theta_i \sigma_i+\beta \sum_{i<j} J_{ij}\sigma_i \sigma_j},
\end{equation}
where $Z[\underline{J},\underline{\theta}]$ is a normalization constant, or partition function. In the memory phase, the system will explore configurations close to a pattern, provided that the initial configuration lies in the basin of attraction of that pattern.
In the following we use the wording patterns, basins of attraction or states equivalently.
In a given state the average activity (or magnetization) and correlations are denoted by $m_i^{\mu}\equiv \langle \sigma_i \rangle_{\mu} $
and $c_{ij}^{\mu}\equiv \langle \sigma_i \sigma_j \rangle_{\mu}$ where the averages are taken with the Gibbs measure inside that state. Informally, 
a Gibbs state corresponds to a stationary state of the system, and so defined by the average macroscopic quantities in that state.

Suppose that starting from some random initial configuration we observe the system for a long time, measuring $M$ configurations.
The standard way to infer interactions couplings $\underline{J}$, and external fields $\underline{\theta}$  is by maximizing the log-likelihood
of $(\underline{J},\underline{\theta})$, given the experimental data \cite{Bialek-arxiv-2006,Monasson-jphysa-2009}, namely
\begin{eqnarray}\label{logL}
\frac{1}{\beta M} \mathcal{L}(\underline{J},\underline{\theta}) =\sum_i \theta_i m_i^{exp}+\sum_{i<j} J_{ij} c_{ij}^{exp}+F[\underline{J},\underline{\theta}],
\end{eqnarray}
where $m_i^{exp}=\frac{1}{M}\sum_{t=1}^M \sigma_i^t$ and $c_{ij}^{exp}=\frac{1}{M}\sum_{t=1}^M \sigma_i^t\sigma_j^t$ are the experimental
magnetizations and correlations and $F=-\frac{1}{\beta}\log Z$ is the free energy.

One can exploit the concavity of the log-likelihood and use a gradient ascent algorithm  to find the unique parameters maximizing the function. However, this needs an efficient algorithm to compute derivatives  of the free energy $F[\underline{J},\underline{\theta}]$, which in general is a difficult task.  A well known technique which can be used for not too big  systems  is of course the Monte Carlo method (see e.g. \cite{Bialek-arxiv-2007}).
Though under certain limiting assumptions,  there exist  good approximation techniques which are efficient, namely mean field, small-correlation and large-field expansions
\cite{Kappen-neurcomp-1998,Tanaka-pre-1998,Mora-thesis-2007,Monasson-jphysa-2009,Monasson-pnas-2009,Roudi-compneuro-2009,Huang-pre-2010} .

In this paper we resort to the mean-field cavity method, or Belief Propagation (BP), to compute the log-likelihood (see e.g. \cite{Yedidia-artif-2003,Frey-ieee-2001,Braunstein-rsa-2005,Mezard-book-2009}).
This technique is closely related to the Thouless-Anderson-Palmer (TAP) approximation in spin glass literature
\cite{TAP-philmag-1977,Virasoro-book-1987}. The approximation is exact on tree graphs and asymptotically correct as long as the graph is locally tree-like or the correlations between variables are sufficiently weak. In spin glass jargon, the approximation works well in the so called replica symmetric phase.

In the BP  approach, the marginals of variables and their joint probability distribution (which is assumed to take a factorized form where only  pair correlations are kept) are estimated by solving a set of  self-consistency functional equations, by exchanging  messages along the edge of the interaction graph (see Ref. \cite{Mezard-book-2009} for comprehensive review).
A message (typically called ``cavity belief'')  $\pi_{i \to j}(\sigma_i)$ is the probability that spin $i$ takes state $\sigma_i$ ignoring the interaction with its neighbor $j$, i.e. in a cavity graph. We call this probability distribution a cavity message. Assuming a tree interaction graph we can write an equation for $\pi_{i \to j}(\sigma_i)$ relating it to other cavity messages $\pi_{k \to i}(\sigma_k)$ sent to $i$: 
\begin{eqnarray}
\pi_{i \to j}(\sigma_i)\propto e^{\beta \theta_i \sigma_i} \prod_{k \in \partial i \setminus j} \left(\sum_{\sigma_j} e^{\beta J_{ij}\sigma_i \sigma_j}\pi_{k \to i}(\sigma_k) \right),
\end{eqnarray}
as in cavity graphs the neighboring variables are independent of each other. These are BP equations and can be used even in loopy graphs to estimate the local marginals.
The equations are solved by starting from random initial values for the cavity messages and updating them in some random sequential order till a fixed point is reached.
Upon convergence the cavity messages are used to obtain the local marginals or ``beliefs'':
\begin{eqnarray}
\pi(\sigma_i,\sigma_j)\propto e^{\beta J_{ij}\sigma_i \sigma_j}\pi_{i \to j}(\sigma_i) \pi_{j \to i}(\sigma_j).
\end{eqnarray}
These marginals are enough to compute the magnetizations $m_i$ and correlations $c_{ij}$ and thus can be used for maximizing the  log-likelihood  by updating the parameters as
\begin{eqnarray}\label{update-J}
\theta_{i}=\theta_{i}+\eta (m_{i}^{exp}-m_{i}), \\ \nonumber
J_{ij}=J_{ij}+\eta (c_{ij}^{exp}-c_{ij}),
\end{eqnarray}
with $\eta\ll 1$ and positive.  Repeating this procedure for sufficient times leads to an estimate of the unknown parameters. Assuming that the  external fields are absent, the inference error can be written as:
\begin{equation}
\Delta_J=\sqrt{\frac{\sum_{i,j\in \partial i}(J_{ij}^{true}-J_{ij})^2}{KN}}.
\end{equation}
A more accurate estimate of the correlations can be obtained by exploiting the Fluctuation-Response theorem $c_{ij}=\partial{m_i}/\partial{\theta_j}$.
This method, called Susceptibility Propagation \cite{Mora-thesis-2007,Mora-jphysio-2009}, uses derivatives of cavity messages (cavity susceptibilities).
Its time complexity grows as $KN^2$, to be compared with  the $K N$ complexity of BP equations.

In this paper we will work exclusively with the BP estimate which is simple and
accurate enough for our studies. Actually, if one is interested only  on
correlations along the edges of a sparse graph, the BP estimation would be as
good as the one obtained by susceptibility propagation. The reader can find more on 
the susceptibility propagation in \cite{Marinari-jstat-2010,Aurell-epjb-2010}.

\section{Inference in the non-ergodic phase}\label{S3}

In an ergodic phase a system visits all configuration space. Sampling for a
long time is well represented by the measure in (\ref{gibbs}).

In a non-ergodic phase, as happens for the Hopfield model in the memory phase,
the steady state of a system is determined by the initial conditions.  Starting
from a configuration close to pattern $\alpha$, the system spends most of its time
(depending on the size of system) in that state. We indicate  the probability
measure which describes such a situation by
$\mathcal{P}_{\alpha}(\underline{\sigma})$, that is the Gibbs measure  restricted
to state $\alpha$. If configurations are sampled from one state, then the expression for the
log-likelihood in (\ref{logL}) should be corrected by replacing $F$ with
$F_{\alpha}$, the free energy of state $\alpha$. Still the log-likelihood is a
concave function of its arguments and so there is a unique solution to this
problem.

It is well known that the Bethe approximation is asymptotically exact in the ergodic phase (\cite{Mezard-book-2009}). 
In this case, the Gibbs weight can be approximately expressed in terms of one- and two- point marginals
$P_{ij}(\sigma_i,\sigma_j)$, $P_i(\sigma_i)$ as follows:
\begin{equation}
\label{eq:pbethe}
\mathcal{P}(\underline{\sigma})\simeq \prod_i P_i(\sigma_i)
	\prod_{i<j}\frac{P_{ij}(\sigma_i,\sigma_j)}{P_i(\sigma_i)P_j(\sigma_j)}.
\end{equation}
The above equation is exact only asymptotically 
(on a replica-symmetric system); it can be used for inference in at least two ways:
The simplest one is by replacing $P_{ij}(\sigma_i,\sigma_j)$ and $P_i(\sigma_i)$  
in the above expression by their experimental estimation (given as input of the inference process), 
equating (\ref{eq:pbethe}) to (\ref{gibbs}) and solving for $\underline{J}$ and $\underline{\theta}$. 
This is known as the ``independent pairs'' approximation. 
A second one, often more precise but computationally more involved, is to search 
for a set of $\underline{J},\underline{\theta}$ and a corresponding 
$\underline{J},\underline{\theta}$-fixed point of BP equations, such that the 
Bethe estimation $\pi_{ij}(\sigma_i,\sigma_j)$, $\pi_i(\sigma_i)$ of 
the two- and one-point marginals match the experimental input as accurately as possible.

In a non-ergodic phase, it is known however that BP equations typically do not converge 
or have multiple fixed points. This is normally attributed to the fact that the BP 
hypothesis of decorrelation of cavity marginals fails to be true. When a BP fixed 
point is attained, it is believed to approximate marginals inside a single state 
(and not the full Gibbs probability), as the decorrelation hypothesis are satisfied 
once statistics are restricted to this state \cite{Mezard-book-2009}.

The fact that BP solutions correspond to restriction to subsets of the original 
measure may suggest that there is little hope in exploiting (\ref{eq:pbethe})
on such systems. Fortunately, 
this is not the case. For every finite system, and every BP fixed point $\alpha$ the following holds,
\begin{equation}
\mathcal{P}(\underline{\sigma})= 
	\frac{Z^{\alpha}_{Bethe}}{Z[\underline{J},\underline{\theta}]} \prod_i \pi^\alpha_i(\sigma_i)
	\prod_{i<j}\frac{\pi^\alpha_{ij}(\sigma_i,\sigma_j)}{\pi^\alpha_i(\sigma_i)\pi^\alpha_j(\sigma_j)}.
\label{eq:pbethe2}
\end{equation}
A proof of a more general statement will be given in appendix \ref{app-BP}. As in the ergodic case,
(\ref{eq:pbethe2}) can be exploited in at least two ways: one is by replacing
$\pi^\alpha_{ij}(\sigma_i,\sigma_j)$ and $\pi^\alpha_i(\sigma_i)$ by their
experimental estimation inside a state and solving for $\underline{J},\underline{\theta}$ the
identity between (\ref{eq:pbethe2}) and (\ref{gibbs}), exactly like in the
independent pairs approximation, as if one just forgets that the samples come
from a single ergodic component. A second one is by inducing BP equations to
converge on fixed points corresponding to appropriate ergodic components. 
In this paper we will take the latter option.

Please notice that the second method, as an algorithm, is more flexible with 
respect to the first one; indeed, there is no reason to have a BP fixed point for 
any set of experimental data, especially when the number of samples is not too large.   
It means that matching exactly the data with those of a BP fixed point is not always possible. Therefore, a better
strategy would be to find a good BP solution which is close enough to the experimental data.

Ignoring the information that our samples come from a
single ergodic component would result in a large inference error due to the
maximization of the wrong likelihood. As an example, we take a tree graph
with Ising spins interacting through random couplings $-1\le J_{ij} \le +1$, in
zero external fields $\theta_i=0$.  Choose an arbitrary pattern $\underline{\xi}$ and
fix a fraction $q$ of the boundary spins to the values in $\underline{\xi}$. 
For $q=0$ the system would be in paramagnetic phase for any finite $\beta$, therefore 
the average overlap of the internal spins with the pattern would be zero. 
On the other hand, for $q=1$ and low temperatures the overlap would be greater than zero,
as expected from a localized Gibbs state around pattern $\underline{\xi}$.  
In this case the observed magnetizations are nonzero and without any information about the
boundary condition we may attribute these magnetizations to external fields which in turn 
result to a large inference error in the couplings.
     
Equivalently, we could put the boundary spins free but restrict the spin configurations 
to a subspace, for instance a sphere of radius $d$ centered
at pattern $\underline{\xi}$ in the configuration space $\Omega_d(\underline{\xi})$.
That is, the system follows the following measure:
\begin{equation} \mathcal{P}_d(\underline{\sigma})\propto
I(\underline{\sigma} \in \Omega_d(\underline{\xi})) e^{\beta \sum_{i<j}
J_{ij}\sigma_i \sigma_j}, 
\end{equation} 
where $I(\underline{\sigma} \in
\Omega_d(\underline{\xi}))$ is an indicator function which selects
configurations in the subspace $\Omega_d(\underline{\xi})$.
By the BP approximation we can compute the magnetizations $m_i^{(d)}$ and the correlations
$c_{ij}^{(d)}$, see Appendix \ref{app-BPd} for more details. Taking these as
experimental data, we may perform the inference by assuming that our data
represent the whole configuration space. This again would result to a large inference error  
(for the same reason mentioned before) whereas taking into account that the system is limited to $\Omega_d(\underline{\xi})$,
we are able to infer the right parameters by maximizing the correct likelihood; i.e. replacing
the total free energy $F$ in the log-likelihood with $F_{\Omega_d(\underline{\xi})}$, the free energy associated to subspace $\Omega_d(\underline{\xi})$.   

\begin{figure}
\includegraphics[width=10cm]{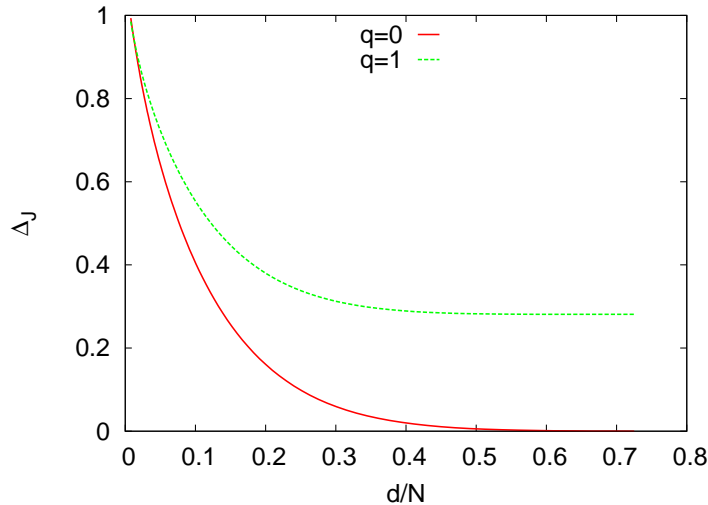}
\caption{ \label{f1}
Inference error on a Cayley tree when the leaves are free ($q=0$) or fixed ($q=1$) to random configuration $\underline{\xi}$. The data come from subspace $\Omega_d(\underline{\xi})$ (a sphere of radius $d$ centered at $\xi$). In the inference algorithm we ignore the boundary condition and that samples are restricted to $\Omega_d(\underline{\xi})$. The internal nodes have degree $K=3$ and size of the tree is $N=766$.}
\end{figure}

In figure $\ref{f1}$ we display the
inference error obtained by ignoring the prior information in the above two cases. Notice that in principle the error would be
zero if we knew $\Omega_d(\underline{\xi})$ and the boundary condition. As it is seen in the figure, the error
remains nonzero when the boundary spins are fixed ($q=1$) even if sampling is performed over the whole space.

\section{Inference in the Hopfield model}\label{S4}

The Hopfield model can be found in three different phases. For large temperatures the system is in the paramagnetic phase where, in the absence of
external fields, magnetizations $m_i$ and overlaps $O^{\mu}=\frac{1}{N}\sum_i \xi_i^{\mu}m_i$ are zero. If the number of patterns is smaller than the critical value $P_c$,
for small temperatures the system enters the memory phase where the overlap between the patterns and the configurations belonging to states selected by the initial conditions can be  nonzero. 
For $P>P_c$,  the Hopfield model at low temperature enters in  a spin glass phase, where the overlaps are typically zero.
In fully connected graphs $P_c \simeq 0.14 N$ and in random Poissonian graphs 
$P_c\simeq 0.637 \langle k \rangle$ where $\langle k \rangle \gg 1$ is the average degree \cite{Coolen-jphysa-2003,Pan-thesis}.

Take the Hopfield model with zero external fields and in the memory phase. We measure samples from a  Glauber dynamics which starts from a configuration close to a pattern $\nu$. The system will stay for a long time in the state $\nu$ and is thus well described by the restricted Gibbs measure  $\mathcal{P}_{\nu}(\underline{\sigma})$. In a configuration $\underline{\sigma}$, the local field seen by  neuron $i$ is   $ h_i=\sum_{j\in\partial i}J_{ij}\sigma_j=\frac{1}{K}\xi_i^{\nu}\sum_{j\in\partial i}\xi_j^{\nu}\sigma_j+\frac{1}{K}
\sum_{\mu\neq\nu}\xi_i^{\mu}\sum_{j\in\partial i}\xi_j^{\mu}\sigma_j$.
If $\nu$ corresponds to the retrieved  pattern, the first term (signal) would have the dominant contribution to $h_i$. The last term (noise)
is a contribution of the other patterns to the local field. To exploit this information, we look for a set of couplings that
result to a Gibbs state equivalent to the observed state of the system. One way to do this is by introducing an auxiliary
external field pointing to the experimental magnetizations, i.e. $\theta_i=\lambda m_i^{exp}$, for a positive $\lambda$;
we may set the couplings $J_{ij}$ at the beginning to zero and compute our estimate of the
correlations $c_{ij}$ by the BP algorithm. This can be used to update the couplings by a small amount
in the direction that maximizes the likelihood, as in (\ref{update-J}) (we do not update the external fields which for simplicity are assumed to be zero).
This updating is repeated iteratively, decreasing $\lambda$ by a small amount in each step. The procedure ends when  $\lambda$ reaches the value zero.

\begin{figure}
\includegraphics[width=10cm]{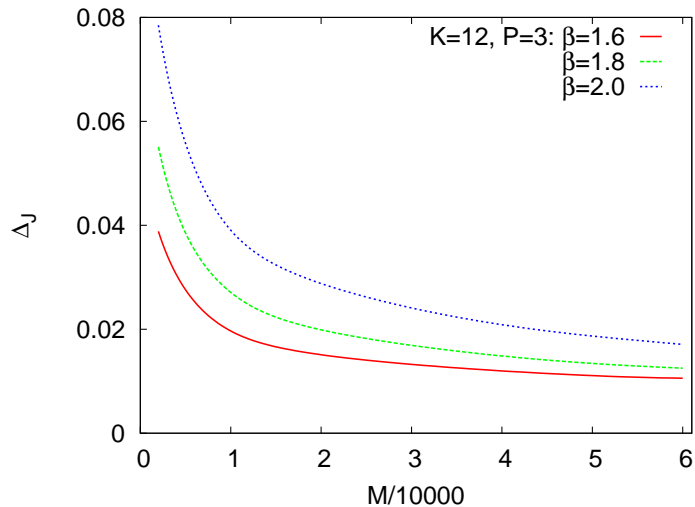}
\caption{ \label{f2}
Inference error versus number of samples for different temperatures. The data are extracted from one pure state of the Hopfield model in the memory phase ($\beta=2$).
Size of the system is $N=1000$, each spin interacts with $K=12$ other randomly selected spins, and number of stored patterns is $P=3$. In the inference algorithm we use $\eta=0.02$.}
\end{figure}

The auxiliary field is introduced only to induce convergence of the equations towards a fixed point giving statistics inside a particular state.
Figure \ref{f2} compares the inference error obtained
with the above procedure for several values of temperature in the memory phase.
For the parameters in the figure, the inferred couplings from one basin were enough to recover the other
two patterns. In the figure we also see how the error decreases by taking larger number of samples from the system.

In general we may have samples from different states of a system. Let us assume
that in the Hopfield model  we stored $P$ patterns by the Hebb rule but the
samples are from $Q$ basins. The estimated correlations $c_{ij}^{\mu}$ in any state $\mu\in
\{1,\ldots,Q\}$ should be as close as possible to the experimental values 
$c_{ij}^{exp,\mu}$.

A natural generalization of the previous algorithm is the following: As before
we introduce external fields $\theta_i^{\mu}=\lambda m_i^{exp,\mu}$ for each
state $\mu$.  At fixed positive $\lambda$ we compute the estimated BP
correlations for different states. Each of these estimations can be used to update the 
couplings as in the single state case. Specifically, this amounts to make a single 
additive update to the couplings by the average vector $\eta\Delta \overline{c}$ given by
$\Delta \overline{c}_{ij}=(\overline{c}_{ij}^{exp}-\overline{c}_{ij})$, 
where $\overline{c}_{ij}^{exp}=\frac{1}{Q}\sum_{\mu=1}^Q c_{ij}^{exp,\mu}$ 
and $\overline{c}_{ij}=\frac{1}{Q}\sum_{\mu=1}^Q c_{ij}^{\mu}$. Indeed, the addends 
of $\Delta \overline{c}$ will be typically linearly independent, so $\Delta \overline{c} = 0$ 
will imply $c_{ij}^{exp,\mu}=c_{ij}^{\mu}$ for $\mu=1,\dots,Q$.

\begin{figure}
\includegraphics[width=10cm]{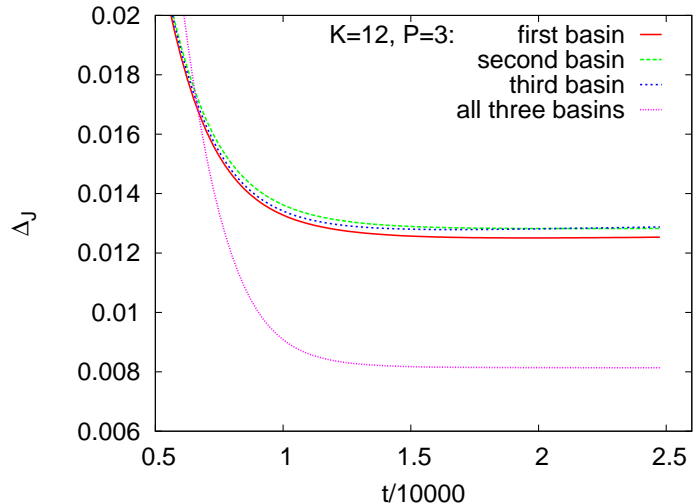}
\caption{ \label{f3}
Evolution of the inference error with update iterations. The data obtained by sampling from one or several pure states of the Hopfield model in the memory phase ($\beta=2$).
The total number of samples in each case is $M=60000$.
Size of the system is $N=1000$, each spin interacts with $K=12$ other randomly selected spins, and number of stored patterns is $P=3$. In the inference algorithm we use $\eta=0.02$.}
\end{figure}

\begin{figure}
\includegraphics[width=10cm]{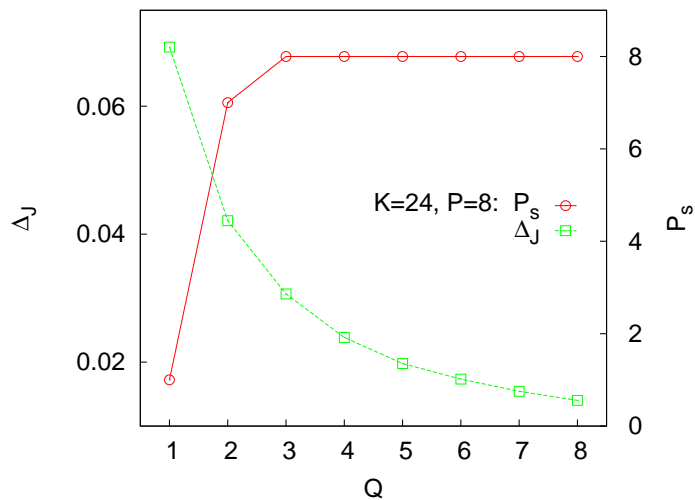}
\caption{ \label{f4}
Inference error and number of states which are stable and highly correlated with the patterns after sampling from $Q$ pure states of the Hopfield model in the memory phase ($\beta=2$).
Size of the system is $N=1000$, each spin interacts with $K=24$ other randomly selected spins, and number of stored patterns is $P=8$. In the inference algorithm we use $\eta=0.02$ and number of samples is $M=Q\times 20000$.}
\end{figure}

We then decrease $\lambda$ and do the BP computation and update steps. Again we
have to repeat these steps until the external field goes to zero.  Figures
\ref{f3} and \ref{f4} show how the inference error changes with sampling from
different states.  Notice that if we had an algorithm that returns exact
correlations, an infinite number of samples from one state would be enough to
infer the right interactions in the thermodynamic limit.  However, given that
we are limited by the number of samples, the learning process is more efficient
if  this number is taken from different states instead of just one.

\section{From inference to unsupervised learning }\label{S5}

Hebbian learning is a stylized way of representing learning processes. Among the many oversimplifications that it involves there is the fact
that patterns are assumed to be presented to the networks through very strong biasing signals.
On the contrary it is of biological interest to consider the opposite limit where only  weak signals are allowed and retrieval takes place with sizable amount of errors.
In  the spin language we are thus interested in the case in which  the system is only slightly biased toward the patterns during the learning phase.
 In what follows we show that one can ``invert'' the inference process discussed in the previous sections and define a local learning rule which copes efficiently with this problem.

As first step we consider a learning protocol in which the patterns are presented sequentially and  in random order to the system by applying an external field in direction of the pattern, that is a field $\theta_i^{\mu}=\lambda \xi_i^{\mu}$ with $\lambda>0$. We assume that initially all couplings are zero. 
Depending on the strength of the field, the system will be forced to explore configurations at different overlaps with the presented pattern $\mu$.
A small $\lambda$ corresponds to a weak or noisy learning whereas for large $\lambda$ the system has to remain
very close to the pattern. What is a small or large $\lambda$, of course depends on the temperature and strength of the couplings. 
Here we assumed the couplings are initially zero, so $\beta \lambda \simeq 1$ defines the boundary between weak and strong fields.

The learning algorithm should indeed force the system to follow a behavior that is suggested by the auxiliary field. 
Therefore, it seems reasonable if we try to match the correlations in the two cases: in absence and presence of the field.   
Notice to the similarities and differences with the first part of the study. As before we are to update the couplings according to deviations in the correlations.
But, here the auxiliary field is necessary for the learning; without that the couplings would not be updated anymore. Moreover, it
is obvious that we can not match exactly the correlations in absence and presence of an external field. We just push the system for a while
towards one of the patterns to reach a stationary state in which all the patterns are remembered.     

For any $\lambda$ we can compute the correlations $c_{ij}^{\lambda,\mu}$ by either sampling from the Glauber dynamics or by
directly running BP, with external fields $\theta_i^{\mu}=\lambda \xi_i^{\mu}$. At the same time we can compute correlations $c_{ij}^{\mu}$ by the BP algorithm in zero external fields and with initial messages corresponding to pattern $\mu$. Then we try to find couplings which match the correlations in the two cases, namely we update the couplings by a quantity  $\eta(c_{ij}^{\lambda,\mu}-c_{ij}^{\mu})$.
The process is repeated  for all couplings and for $t_L =O(1)$ iterations with the same pattern $\mu$. Next we switch to some other randomly selected pattern $\nu$ and the whole process is repeated for $T_L$ learning steps. Notice that here $\lambda$ is fixed from the beginning.

\begin{figure}
\includegraphics[width=10cm]{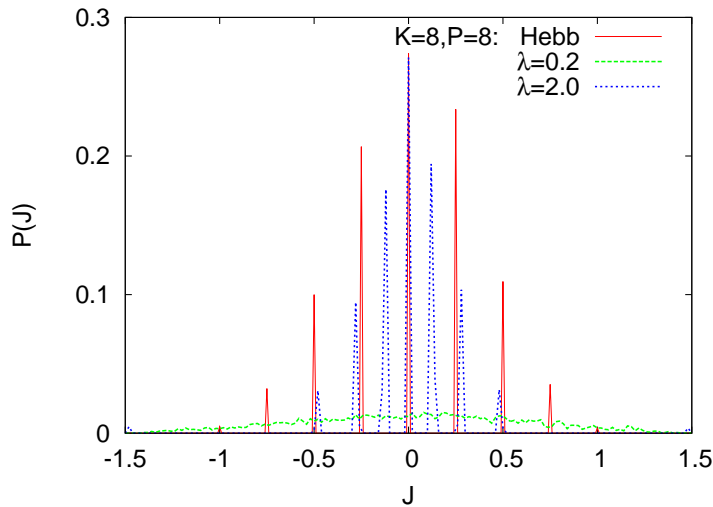}
\caption{ \label{f5}
Comparing the histogram of Hebbian couplings with that of learned couplings for small and large values of the external field. 
Size of the system is $N=1000$, each spin interacts with $K=8$ other randomly selected spins. Here we are to store $P=8$ random and uncorrelated patterns.
In the learning algorithm we use $\beta=2$ and $\eta=0.003$.}
\end{figure}

The above learning protocol displays a range of interesting phenomena. Firstly one notices  that
for $c_{ij}^{\lambda,\mu}\simeq \xi_i^{\mu} \xi_j^{\mu}$ (i.e. for very large external fields) and $c_{ij}^{\mu}\simeq 0$ (i.e. for very high temperature or isolated neurons) the above learning results to the Hebb couplings of the Hopfield model. In Figure \ref{f5} we compare the histogram of learned couplings for small and large $\lambda$ with the Hebbian ones.

\begin{figure}
\includegraphics[width=10cm]{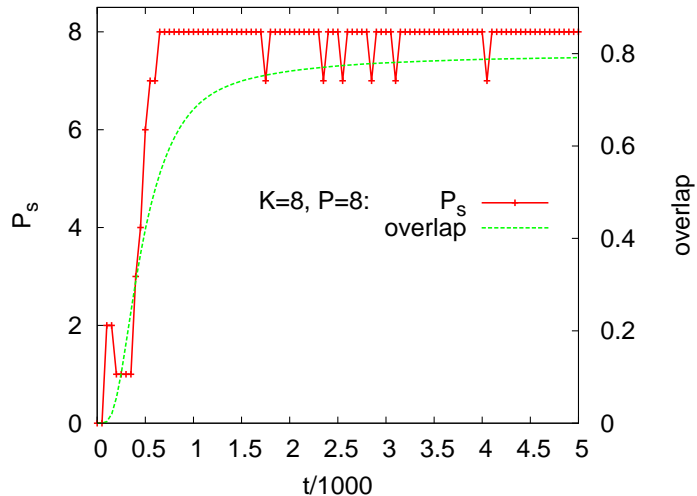}
\caption{ \label{f6}
Evolution of the average overlap and fraction of successfully learned patterns in the learning algorithm.
Size of the system is $N=1000$, each spin interacts with $K=8$ other randomly selected spins. In the learning algorithm we set $\lambda=0.2$,  $\beta=2$, $\eta=0.003$,
and number of patterns that are to store is $P=8$.}
\end{figure}

\begin{figure}
\includegraphics[width=10cm]{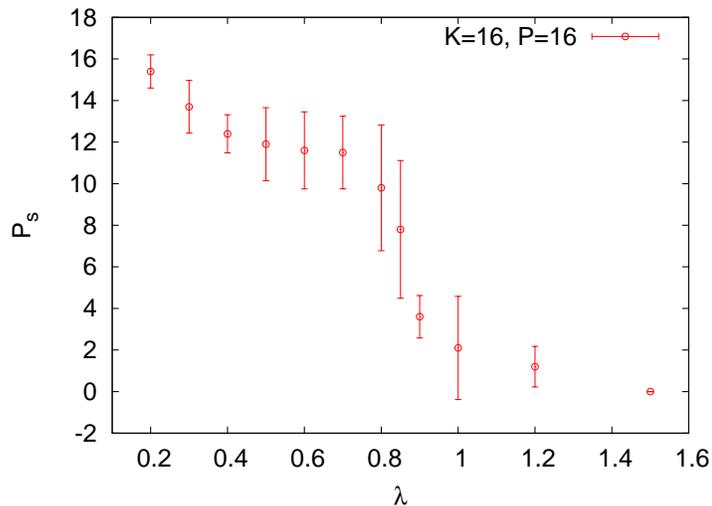}
\caption{ \label{f7}
Average number of successfully learned patterns in the learning algorithm for different values of $\lambda$.
Size of the system is $N=1000$, each spin interacts with $K=16$ other randomly selected spins. 
The learning algorithm works at $\beta=2, \eta=0.003$, and number of patterns that are to store is $P=16$. The average is taken over $10$ realizations of the patterns.}
\end{figure}

The number of patterns $P_s$ which are highly correlated with stable configurations depends on the strength of external fields. 
We consider that pattern $\mu$ is ``learned'' if there is a Gibbs state with nonzero overlap $O^{\mu}$ that is definitely larger than the other ones $\{O^{\nu}| \nu\ne \mu\}$.
Figures \ref{f6} and \ref{f7} show how these quantities evolve during the learning process and by increasing the magnitude of external filed.

\begin{figure}
\includegraphics[width=10cm]{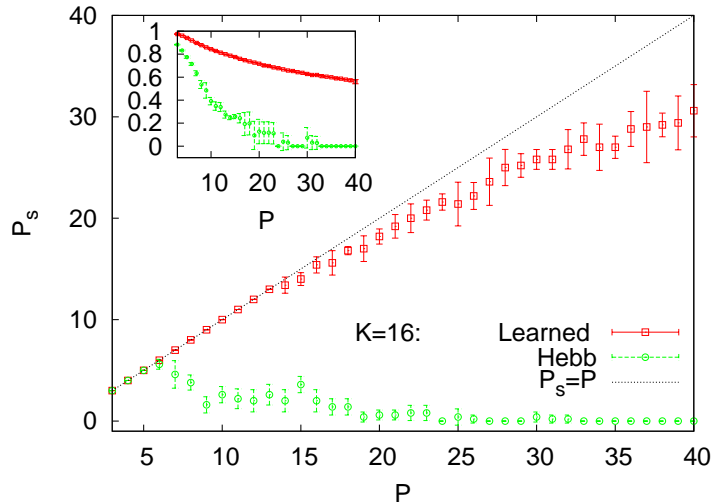}
\caption{ \label{f8}
Average number of successfully learned patterns in the learning algorithm and Hebb rule versus $P$, the number of patterns that are to store. 
The inset shows the average overlap.
Size of the system is $N=1000$, each spin interacts with $K=16$ other randomly selected spins. The learning algorithm works at $\lambda=0.2, \beta=2, \eta=0.003$.
The average is taken over $10$ realizations of the patterns.}
\end{figure}

For small $P$ nearly all patterns are learned, whereas, for larger $P$ some patterns are missing.
A large number of patters can thus be learned at the price of  smaller overlaps and weaker states. That is, the average
overlap in successfully learned patterns decreases continuously by increasing $P$, approaching the paramagnetic limit.
In Figure \ref{f8} we compare this behavior with that of Hebb couplings.

As the figure shows, there is a main qualitative difference between Hebbian learning of the Hofield model and the protocol discussed here.
In the former case when the number of stored patterns exceeds some critical value the systems enters in a spin glass phase where all memories are lost and the BP algorithm does not converge anymore. On the contrary, in our case  many patterns can be stored without ever entering the spin glass phase (for a wide range of choices of $\lambda$). The BP algorithm always converges, possibly to a wrong fixed point if the corresponding pattern is not stored.

\subsection{Population dynamics analysis of the learning protocol}\label{S5-1}

\begin{figure}
\includegraphics[width=10cm]{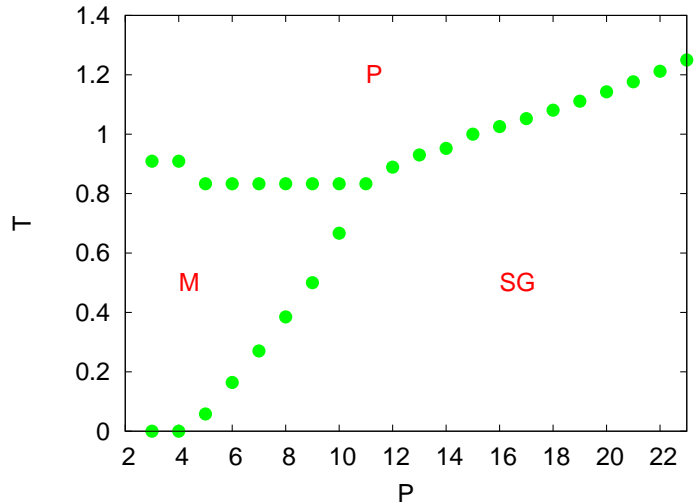}
\caption{ \label{f9}
The phase diagram of Hopfield model on random regular graphs of degree $K=12$ obtained with population dynamics ($N_p=10^5$).
Horizontal axes is number of patterns $P$ and vertical axes is temperature $T=1/\beta$. The paramagnetic, memory and spin glass phases are labeled with $P, M$ and $SG$, respectively.}
\end{figure}

Population dynamics is usually used to obtain the asymptotic and average  behavior of quantities that obey a set of deterministic or stochastic equations \cite{Mezard-physjb-2001}.  For instance, to obtain the phase diagram of the Hopfield model with population dynamics one introduces a population of $N_P$ messages representing the BP cavity messages in a reference state, e.g. pattern $\underline{\xi}^{\nu}=+1$ \cite{Pan-thesis}. Then one updates the messages in the population according to the BP equations: at each time step a randomly selected cavity message is replaced with a new one computed by $K-1$ randomly selected ones appearing on the r.h.s. of the BP equations.
In each  update, one generates the random couplings $J_{ij}=\frac{1}{K}+\frac{1}{K}\sum_{\mu\ne \nu}\xi_i^{\mu}\xi_j^{\mu}$ by sampling
the other $P-1$ random patterns.  After a sufficiently large number of updates, one can compute the average overlap with the condensed
pattern $\nu$ to check if the system is in a memory phase. The stability of condensed state would depend on the stability of the above
dynamics with respect to small noises in the cavity messages. If $N_P$ is large enough, one obtains the phase diagram
of Hopfield model in the thermodynamic limit averaged over the ensemble of random regular graphs and patterns.
We used the above population dynamics to obtain the phase diagram of the Hopfield model on random regular graphs, see Figure \ref{f9}.

\begin{figure}
\includegraphics[width=10cm]{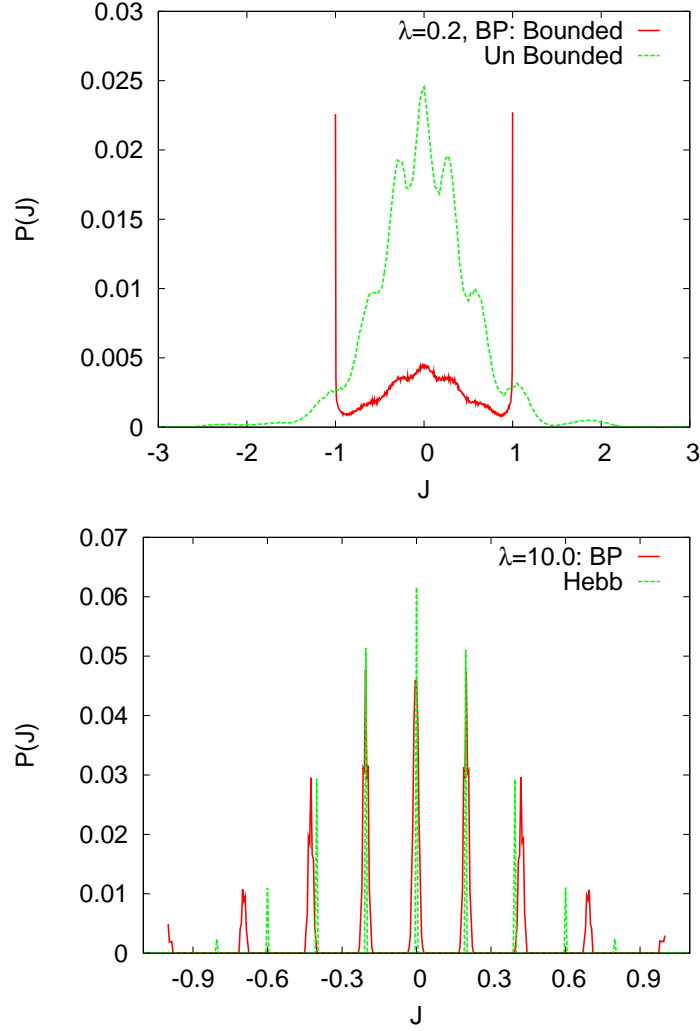}
\caption{ \label{f10}
The histogram of learned couplings obtained by the population dynamics in random regular graphs of degree $K=8$. Number of patterns that are to store is $P=8$. 
In the upper panel we compare the two cases of learning with bounded and unbounded couplings for a small external field. In the lower panel we compare the Hebb rule with the learning algorithm for a large 
external field. In the algorithm we use $N_p=1000$, $\beta=1$, and $\eta=0.01$.}
\end{figure}

In order to study the new learning protocol we need a more sophisticated population dynamics. The reason is that in contrast to Hebb couplings,
we do not know in advance the learned couplings.
In Appendix \ref{app-pop}   we explain in more details the population dynamics that we use to analyze the learning process studied in this paper.
The algorithm is based on $P$ populations of BP messages and one population of couplings. These populations
represent the probability distributions of BP messages in different states and couplings over the interaction graph.
For a fixed set of patterns $\{\underline{\xi}^{\mu}|\mu=1,\ldots,P\}$ we update the populations according to the BP equations
and the learning rule, to reach a steady state. Figure \ref{f10} displays the histogram of couplings obtained in this way.
In the figure we compare two cases of bounded and unbounded couplings. In the first case the couplings should have a magnitude less
than or equal to $1$ whereas in the second case they are free to take larger values. We observe a clear difference between the two
cases; when $\lambda$ is small, the couplings are nearly clipped in the bounded case whereas the unbounded couplings go beyond $\pm 1$. However, in both cases there is some structure in the range of small couplings. Increasing the magnitude of $\lambda$ we get more and more structured couplings.
For very large fields they are similar to the Hebb couplings. For small fields the histogram of the couplings
is very different from the Hebb one,  though the sign of the learned and the Hebbian couplings is the same.

There are a few comments to mention here; in the population dynamics we do not have a fixed graph structure and to
distinguish $P$ patterns from each other we have to fix them at the beginning of the algorithm. Moreover, we have to modify
the BP equations to ensure that populations are representing the given patterns, see Appendix \ref{app-pop}.
And finally the outcome would be an average over the ensemble of random regular graphs, for a fixed set of patterns.

Having the stationary population of couplings, one can check the stability of each state by
checking the stability of the BP equations at the corresponding fixed point. The maximum capacity that we obtain in this way for the learned couplings
is the same as the Hebb one whereas on single instances we could store much larger number of patterns.
The reason why we do not observe this phenomenon in the population dynamics resides in the way that we are checking stability;
the fixed patterns should be stable in the ensemble of random regular graphs.
In other words, checking for stability in the population dynamics is stronger than checking it in a specific graph.

The main result of our analysis consists in showing that the distribution of the couplings arising from the BP learning protocol
is definitely different from the Hebbian one.

\section{Discussion and perspectives}\label{S6}

We studied the finite connectivity inverse Hopfield problem at low temperature, where the data are sampled from a non-ergodic regime.
We showed that the information contained in the  fluctuations within single pure states can be used to infer the correct interactions.

We also used these  findings to design a simple learning protocol  which is able to store patterns learned under noisy conditions.
Surprisingly enough it was possible to show that  by demanding a small though finite overlap with the patterns it is possible to store a
large number of patterns without ever reaching a spin glass phase. The learning process avoids the spin glass phase by decreasing
the overlaps, as the number of patterns increases.
A separate analysis which is similar to the one presented in Ref. \cite{Baldassi-pnas-2007} (and not reported here) shows that the equations can be heavily simplified
without loosing their main learning capabilities.

In this paper we focused on a simple model of neural networks with symmetric couplings.
It would be interesting to study more realistic models like the integrate and fire model of neurons with
general asymmetric couplings.  Moreover, instead of random unbiased patterns one may consider
sparse patterns which are more relevant in the realm of neural networks.

The arguments presented in this paper can also be relevant to problem of inferring a dynamical model for a system by observing its dynamics.
In this case, a system is defined solely by its evolution equations and one cannot rely on the Boltzmann equilibrium distribution.
Still it is possible to try to infer the model by writing the likelihood for the model parameters given the data and given the underlying dynamical stochastic process.
A mean-field approach has been recently described in \cite{Roudi-arxiv-2010}. We actually checked this approach in our problem and observed qualitatively the same behavior as the
static approach.  In fact, which method is best heavily depends on the type of data which are available.

\section{Acknowledgements}

The work of  was partially supported by a {\it Programma Neuroscienze} grant by the Compagnia di San Paolo and  the EC grant 265496. 

\appendix

\section{Proof of Exactness of the Bethe expression for arbitrary BP fixed points}\label{app-BP}

A $\beta\to\infty$ limit version of this result (except the determination of the value of the constant $Z_{Gibbs}/Z_{Bethe}$) appeared in \cite{kolmogorov2006convergent}. This result is valid for general (non-zero) interactions. For a family of ``potentials'' $\Psi_a:\underline{x}_a\mapsto \Psi_a(\underline{x}_a)>0$, where we denote by $\underline{x}_a$ the subvector of $\underline{x}$ given by $\{x_i:i\in \partial a\}$. We will use the shorthand $i\in a$ or $a\in i$ to mean $i\in \partial a$.

\emph{Proposition}. Given a factorized probability function
\[
\mathcal{P}\left(\underline{x}\right)=\frac{1}{Z_{Gibbs}}\prod_{a}\Psi_{a}\left(\underline{x}_{a}\right)
\]
and a BP fixed point $\left\{ b_{ia}\right\} _{i\in a,a\in A}$ and plaquette marginals $b_{a}\left(x_{a}\right)=z_{a}^{-1}\Psi_{a}\left(\underline{x}_{a}\right)\prod_{i\in a}b_{ia}\left(x_{i}\right)$
and single marginals $b_{i}\left(x_{i}\right)=\sum_{\left\{ x_{j}:j\in a\setminus i\right\} }b_{a}\left(\underline{x}_{a}\right)$
for every $a\in i$, then
\[
\mathcal{P}\left(\underline{x}\right)=\frac{Z_{Bethe}}{Z_{Gibbs}}\prod_{a}\frac{b_{a}\left(\underline{x}_{a}\right)}{\prod_{i\in a} b_{i}\left(x_{i}\right)}\prod_{i}b_{i}\left(x_{i}\right)\]

\emph{Proof}. Using the fact that $b_{i}\left(x_{i}\right)=z_{i}^{-1}\prod_{a\in i}b_{ai}\left(x_{i}\right)$
and $b_{ia}\left(x_{i}\right)=z_{ia}^{-1}\prod_{e\in i\setminus a}b_{ei}\left(x_{i}\right)$,
we obtain $b_{ia}(x_i)=b_i(x_i) z_i b_{ai}^{-1}(x_i)z_{ia}^{-1}$, then using the definitions: 

\begin{eqnarray*}
\prod_{a}\frac{b_{a}\left(\underline{x}_{a}\right)}{\prod_{i\in a}b_{i}\left(x_{i}\right)}\prod_{i}b_{i}\left(x_{i}\right) & = & \prod_{a}\frac{\Psi_{a}\left(\underline{x}_{a}\right)z_{a}^{-1}\prod_{i\in a}b_{ia}\left(x_{i}\right)}{\prod_{i\in a}z_{ia}z_{i}^{-1}b_{ia}\left(x_{i}\right)b_{ai}\left(x_{i}\right)}\prod_{i}b_{i}\left(x_{i}\right)\\
 & = & \prod_{a}\frac{\Psi_{a}\left(\underline{x}_{a}\right)z_{a}^{-1}}{\prod_{i\in a}z_{ia}z_{i}^{-1}b_{ai}\left(x_{i}\right)}\prod_{i}\frac{\prod_{a\in i}b_{ai}\left(x_{i}\right)}{z_{i}}\\
  & = & Z_{Gibbs}\mathcal{P}\left(\underline{x}\right)\prod_{a}\frac{z_{a}^{-1}}{\prod_{i\in a}z_{ia}z_{i}^{-1}}\prod_{i}z_{i}^{-1}\\
   & = & Z_{Gibbs}\mathcal{P}\left(\underline{x}\right)Z_{Bethe}^{-1}\end{eqnarray*}

This proves that a fixed point can be interpreted as a form of
reparametrization of the original potentials. In fact, a sort of converse also holds:

\emph{Proposition}: If $\mathcal{P}(\underline x)\propto \prod_a \Psi_a(\underline x_a)$ satisfies a Bethe-type expression 

\begin{equation}
\mathcal{P}\left(\underline{x}\right)\propto\prod_{a}\frac{b_{a}\left(\underline{x}_{a}\right)}{\prod_{i\in a} b_{i}\left(x_{i}\right)}\prod_{i}b_{i}\left(x_{i}\right)
\label{eq:bethe3}
\end{equation}
with $\sum_{\left\{ x_{j}:j\in a\setminus i\right\} }b_{a}\left(\underline{x}_{a}\right)=b_{i}\left(x_{i}\right)$ for every $i\in a$.
Then there exists a BP fixed point $\left\{ b_{ia}\right\} _{i\in a,a\in A}$
such that $b_{a}\left(\underline{x}_{a}\right)\propto\Psi_{a}\left(\underline{x}_{a}\right)\prod_{i\in a}b_{ia}\left(x_{i}\right)$. 

\emph{Proof}: Choose any configuration $\underline{z}$. We will use the following notation: $\underline{z}_{a\setminus i}=\{z_j\}_{j\in a\setminus i}$, and $\underline{z}_{-i}=\{z_j\}_{j\neq i}$. Define $b_{ia}\left(x_{i}\right)\propto\Psi_{a}^{-1}\left(x_{i},\underline{z}_{a\setminus i}\right)b_{a}\left(x_{i},\underline{z}_{a\setminus i}\right)$,
normalized appropriately. Afterwards, we can define $b_{ai}\left(x_{i}\right)\propto b_{i}\left(x_{i}\right)b_{ia}^{-1}\left(x_{i}\right)$.

By definition of
$\mathcal{P}$ we have $\mathcal{P}\left(x_{i}|\underline{z}_{-i}\right)\propto\prod_{a\in i}\Psi_{a}\left(x_{i},\underline{z}_{a\setminus i}\right)$.
Similarly, but using (\ref{eq:bethe3}), and noting by $n_i=|\partial i|$, we have also 
$\mathcal{P}\left(x_{i}|\underline{z}_{-i}\right)\propto b_{i}\left(x_{i}\right)^{1-n_{i}}\prod_{a\in i}b_{a}\left(x_{i},\underline{z}_{a\setminus i}\right)$.
Then $b_{i}\left(x_{i}\right)^{n_{i}-1}\propto\prod_{a\in i}\frac{b_{a}\left(x_{i},\underline{z}_{a\setminus i}\right)}{\Psi_{a}\left(x_{i},\underline{z}_{a\setminus i}\right)}\propto\prod_{a\in i}b_{ia}\left(x_{i}\right)\propto\prod_{a\in i}b_{i}\left(x_{i}\right)\prod_{a\in i}b_{ai}^{-1}\left(x_{i}\right)$,
and thus $b_{i}\left(x_{i}\right)\propto\prod_{a\in i}b_{ai}\left(x_{i}\right)$. This also implies that $b_{ia}(x_i)\propto b_i(x_i) b_{ai}^{-1}(x_i) \propto \prod_{e\in i\setminus a} b_{ei}(x_i)$, proving that the first BP equation is satisfied.

By definition of $\mathcal{P}$, $\mathcal{P}\left(\underline{x}_{a}|\underline{z}_{-a}\right)\propto\Psi_{a}\left(\underline{x}_{a}\right)\prod_{i\in a}c_{i}\left(x_{i}\right)$
where $c_{i}\left(x_{i}\right)=\prod_{e\in i\setminus a}\Psi_{e}\left(x_{i},\underline{z}_{e\setminus i}\right)$.
Moreover using (\ref{eq:bethe3}), we can
conclude that also $\mathcal{P}\left(\underline{x}_{a}|\underline{z}_{-a}\right)\propto b_{a}\left(\underline{x}_{a}\right)\prod_{i\in a}d_{i}\left(x_{i}\right)$
where $d_{i}\left(x_{i}\right)=b_{i}\left(x_{i}\right)^{1-n_{i}}\prod_{e\in i\setminus a}b_{e}\left(x_{i},\underline{z}_{e\setminus i}\right).$
This implies that $\frac{b_{a}\left(\underline{x}_{a}\right)}{\Psi_{a}\left(\underline{x}_{a}\right)}\propto\prod_{i\in a}\frac{c_{i}\left(x_{i}\right)}{d_{i}\left(x_{i}\right)}$.
But we also have that $\frac{c_{i}\left(x_{i}\right)}{d_{i}\left(x_{i}\right)}\propto b_{i}\left(x_{i}\right)^{n_{i}-1}\prod_{e\in i\setminus a}b_{ie}^{-1}\left(x_{i}\right)\propto \prod_{e\in i\setminus a}b_{ei}\left(x_{i}\right)\propto b_{ia}\left(x_{i}\right)$,
so $b_{a}\left(\underline{x}_{a}\right)\propto\Psi_{a}\left(\underline{x}_{a}\right)\prod_{i\in a}b_{ia}\left(x_{i}\right)$ as desired. Now $b_i(x_i)=\sum_{\underline{x}_{a\setminus i}}b_a(\underline{x}_a)$ by hypothesis, so $b_{ai}(x_i)\propto b_i(x_i) b_{ia}^{-1}(x_i) \propto \sum_{\underline{x}_{a\setminus i}}\Psi_{a}\left(\underline{x}_{a}\right)\prod_{j\in a\setminus i}b_{ja}\left(x_{j}\right)$ and this proves that the second BP equation is also satisfied.

\section{Computing thermodynamic quantities in a restricted space}\label{app-BPd}

Consider the Ising model on a tree graph of size $N$ with couplings $\underline{J}$ and external fields $\underline{\theta}$.
Suppose that we are given a reference point $\underline{\xi}$ in the configuration space $\{-1,+1\}^N$ and the following measure
\begin{equation}
\mathcal{P}_d(\underline{\sigma})\propto  I(\underline{\sigma}
\in \Omega_d(\underline{\xi})) e^{\sum_i \beta \theta_i \sigma_i+\sum_{i<j} \beta J_{ij}\sigma_i\sigma_j},
\end{equation}
where $\Omega_d(\underline{\xi})$ is a sphere of radius $d$ centered at $\underline{\xi}$.
By distance of two configurations we mean the Hamming distance, i.e. number of spins which
are different in the two configurations. The aim is to compute thermodynamic quantities like average
magnetizations and correlations in an efficient way. We do this by means of the Bethe approximation and so BP algorithm.

First we express the global constraint $I(\underline{\sigma}\in \Omega_d(\underline{\xi}))$ as a set of local constraints by introducing messages $d_{i\to j}$
that each node sends for its neighbors. For a given configuration $\underline{\sigma}$,
\begin{equation}
d_{i\to j}=\sum_{k\in \partial i\setminus j} d_{k\to i}+(1-\delta_{\sigma_i,\xi_i}),
\end{equation}
denotes the distance of $\underline{\sigma}$ from $\underline{\xi}$ in the cavity graph $\mathcal{G}_{i\to j}$ which includes $i$ and all nodes connected to $j$ through $i$.
With these new variables we can write BP equations as
\begin{equation}
\pi_{i\to j}(\sigma_i,d_{i\to j};\sigma_j,d_{j\to i})\propto e^{\beta \theta_i \sigma_i+\beta J_{ij}\sigma_i\sigma_j}\sum_{\{\sigma_k,d_{k\to i}|k\in \partial i\setminus j\}}
 I_i \prod_{k\in \partial i\setminus j}\pi_{k\to i}(\sigma_k,d_{k\to i};\sigma_i,d_{i\to k}),
\end{equation}
where $I_i$ is an indicator function to check the constrains on $d_{i\to k}$ and $\sum_{k\in \partial i}d_{k \to i}+(1-\delta_{\sigma_i,\xi_i})\le d$.
Starting from random initial values for the BP messages we update them according to the above equation. After convergence the local marginals read
\begin{equation}
\pi(\sigma_i,\sigma_j)\propto e^{-\beta J_{ij}\sigma_i\sigma_j}\sum_{d_{i\to j},d_{j\to i}}
 I_{ij} \pi_{i\to j}(\sigma_i,d_{i\to j};\sigma_j,d_{j\to i})\pi_{j\to i}(\sigma_j,d_{j\to i};\sigma_i,d_{i\to j}),
\end{equation}
where in $I_{ij}$ we check if $d_{i\to j}+d_{j\to i}\le d$. These marginals will be used to compute the average magnetizations and correlations.
Notice that when the graph is not a tree we need to pass the messages $d_{i\to j}$ only along the edges of a spanning tree (or chain) which is selected and fixed at the beginning of the algorithm.

\section{Population dynamics}\label{app-pop}

Consider $P$ patterns $\xi_a^{\mu}\in \{-1,+1\}$, where  $\mu=1,\ldots,P$ and
$a$ goes from $1$ to $N_p$, which is equivalent to the size of system. The patterns, learning rate
$\eta$ and parameter $\lambda$ are fixed at the beginning of the algorithm.
To each patten we assign a population of messages $\pi_{a,l}^{\mu}(\sigma)$ where $l=1\ldots,K$ ($K$ is the node degree). These are to represent
the normalized BP messages that we use in the learning algorithm. Besides this we have also a population of couplings $J_{ab}$.

The population dynamics has two update steps: updating the $P$ populations of messages and updating the population of couplings.

To update the messages in population $\mu$ we do the following:

i) select randomly $(a_0,l_0)$ and $\{(a_1,l_1),\ldots,(a_{K-1},l_{K-1})\}$,

ii) use messages $\{\pi_{a_1,l_1}^{\mu},\ldots,\pi_{a_{K-1},l_{K-1}}^{\mu}\}$ and couplings $\{J_{a_0,a_1},\ldots,J_{a_0,a_{K-1}}\}$ to compute a new BP message $\pi_{new}$,

iii) replace message $\pi_{a_0,l_0}^{\mu}(\xi_{a_0}^{\mu})$ with $\max(\pi_{new}(\xi_{a_0}^{\mu}),\pi_{new}(-\xi_{a_0}^{\mu}))$,

Notice to the maximum we are taking in the last step. This is to ensure that BP messages in population $\mu$ are related to pattern $\xi_a^{\mu}$.
We do these updates for $t_{BP}$ iterations, where in each iteration all members of a population are updated in a random sequential way.

To update the couplings we go through the $P$ populations and do the following:

i) select randomly $(a,l_a)$ and $(b,l_b)$,

ii) use messages $\pi_{a,l_a}^{\mu}, \pi_{b,l_b}^{\mu}$ and coupling $J_{ab}$ to compute correlation $c_{ab}^{\lambda,\mu}$,
 i.e. in presence of external fields $\lambda \xi_a^{\mu}$ and $\lambda \xi_b^{\mu}$,

iii) use messages $\pi_{a,l_a}^{\mu}, \pi_{b,l_b}^{\mu}$ and coupling $J_{ab}$ to compute correlation $c_{ab}^{\mu}$, i.e. in absence of the external fields,

iv) update the coupling as $J_{ab}=J_{ab}+\eta (c_{ab}^{\lambda,\mu}-c_{ab}^{\mu})$

The learning updates are done for $t_{L}$ iterations.

All together the population dynamics will have $T_L$ learning steps each one consist of $Pt_{BP}+Pt_{L}$ update iterations.
In practice we set $t_{BP}\simeq 10$ and $t_{L}\simeq 1$.

\end{document}